\documentclass[twoside,twocolumn]{article}

\usepackage{textcomp}

\usepackage{xspace}

\usepackage{gensymb}
\usepackage{float}
\usepackage{multicol}
\usepackage{physics}
\usepackage{siunitx}
\usepackage{amsfonts}
\usepackage{amssymb}
\usepackage{verbatim}
\usepackage{hyperref}
\usepackage{url} 
\usepackage{graphicx}
\usepackage{mathptmx}
\usepackage{amsmath}
\usepackage[citestyle=numeric,bibstyle=numeric, sorting=none, natbib=true, backend=bibtex,hyperref=true,maxbibnames=5,maxcitenames=2]{biblatex}
\usepackage{blindtext} % Package to generate dummy text throughout this template 

\usepackage[english]{babel} % Language hyphenation and typographical rules

\usepackage[hmarginratio=1:1, left = 18mm, top=32mm,columnsep=20pt]{geometry} % Document margins
\usepackage[hang, small,labelfont=bf,up,textfont=it,up]{caption} % Custom captions under/above floats in tables or figures
\usepackage{booktabs} % Horizontal rules in tables

\usepackage{abstract} % Allows abstract customization
\usepackage{titlesec} % Allows customization of titles
\renewcommand\thesection{\Roman{section}} % Roman numerals for the sections
\renewcommand\thesubsection{\roman{subsection}} % roman numerals for subsections
\titleformat{\section}[block]{\large\scshape\centering}{\thesection.}{1em}{} % Change the look of the section titles
\titleformat{\subsection}[block]{\large}{\thesubsection.}{1em}{} % Change the look of the section titles

\usepackage{titling} % Customizing the title section

\usepackage{hyperref} % For hyperlinks in the PDF

\usepackage{soul}   % provides \hl{} for highlighting

\title{Implementation of a generalized precession parameter in the RIFT parameter estimation algorithm} % Article title
\author{%
\textsc{Chad Henshaw$^1$, Richard O'Shaughnessy$^2$, Laura Cadonati$^1$} \\[1ex] % Your name
\normalsize $^1$\emph{School of Physics, Georgia Institute of Technology, Atlanta, GA 30332, USA} \\
\normalsize$^2$\emph{Rochester Institute of Technology, Rochester, NY 14623, USA}% Your institution
\normalsize %\href{mailto:john@smith.com}{john@smith.com} % Your email address
}
\date{\today} % Leave empty to omit a date
\renewcommand{\maketitlehookd}{%
\begin{abstract}
\noindent Since the initial discovery of gravitational waves in 2015, significant developments have been made towards waveform interpretation and estimation of compact binary source parameters. We present herein an implementation  of the generalized precession parameter $\langle \chi_p \rangle$, which averages over all angular variations on the precession timescale, within the RIFT parameter estimation framework.  Relative to the originally-proposed precession parameter $\chi_p$, which characterizes the single largest dynamical spin in a binary, this new parameter $\langle \chi_p \rangle$ has a unique domain $1 < \langle \chi_p \rangle < 2$, which is exclusive to binaries with two precessing spins.   After reviewing the physical differences between these two parameters, we describe how $\langle \chi_p \rangle$ was implemented in RIFT and apply it to all 36 events from the second half of the Advanced LIGO and Advanced Virgo third operating run (O3b).  In O3b,  ten events show significant amounts of precession $\langle \chi_p \rangle > 0.5$. Of particular interest is GW191109\_010717; we show it has a $\sim28\%$ probability that the originating system necessarily contains two misaligned spins.   % Dummy abstract text - replace \blindtext with your abstract text

\end{abstract}
}

\begin{document}

% Print the title
\maketitle

%----------------------------------------------------------------------------------------
%	ARTICLE CONTENTS
%----------------------------------------------------------------------------------------

\section{Introduction}

In the modern era of gravitational wave astronomy, data obtained by the Advanced LIGO and Advanced Virgo detectors offers fresh insights on the characteristics of coalescing binary black holes (BBHs) \cite{gwtc1, gwtc2, gwtc2.1, gwtc3}. Analysis of these data has afforded us a new window on the dynamics of such astrophysical systems, and we are now able to identify the detailed characteristics of their evolution.
%The coalescence of massive binaries produces radiation in the form of gravitational waves, ripples that alternatively squeeze and stretch spacetime as they propagate through it. %After traversing vast distances, these ripples reach and pass through the Earth, subtly distorting the matter that surrounds us. Through the use of Michelson interferometry, gravitational wave detectors such as LIGO can measure these distortions and characterize the gravitational wave signal.\\

%\LC{I am not sure this is the correct introduction - is the increased expected event rate what drives this work? I would instead just put the current catalog in the context of what is to be expected in the future (cite observing scenario paper), then say something about matched filtering and parameter estimation}
%CH: reworded these two paragraphs 11-16
%
%At the time of this publication LIGO is going through a maintenance period, and gearing up for its fourth observing run (O4). 
To date a total of 90 gravitational wave events have been observed \cite{gwtc3}, and it is predicted that in O4 the LIGO-Virgo-KAGRA (LVK) network will detect $10^{+52}_{-10}$ binary neutron star (BNS) mergers, $1^{+91}_{-1}$ neutron star-black hole (NSBH) mergers, and $79^{+89}_{-44}$ BBH mergers \cite{Acernese2018, Aasi2015}. 
%Accordingly, the basis of gravitational wave data collected by the LIGO Scientific Collaboration (LSC) is constantly growing, and our ability to divine and analyze the information that these data contain must be developed in turn. The data recorded by LIGO are first put through a match filter, which rapidly assesses whether those data are consistent with gravitational wave signals and what classification (BBH, BNS, NSBH) they fall into. These raw signal data are then put through a pipeline of post-analyses, including parameter estimation. 
By comparing the recorded signal to that produced by simulated waveforms, we can obtain statistical information about the fundamental source parameters of the binary system. The collaboration currently uses a number of parameter estimation programs, including LALInference \cite{Veitch_2015}, Bilby \cite{Ashton_2019}, and RIFT \cite{lange2018rapid, Rift_algorithm}.
%\LC{how do these codes fundamentally differ from each other? This is the place where to describe what RIFT is (move here some material from the abstract)}\\
These codes
construct posterior probability distributions for the binary’s intrinsic parameters $\left(m_1, m_2, \mathbf{\chi}_1, \mathbf{\chi}_2\right)$ which may then be procedurally transformed into other characteristic parameters to assess the properties of a binary. Such parameterizations are useful both as a comparative heuristic for analysis, and also as a sampling coordinate for improved parameter estimation. One of the properties of a binary system that can be effectively reduced to a parameter is its precession - the change in direction of the binary's orbital angular momentum over time.

The characteristics of a gravitational waveform will be distinctly altered if the source binary contains objects with misaligned spins - i.e. the spin angular momenta of the individual objects do not point in the same direction as the orbital angular momentum of the binary. In such systems the orbital plane will precess about the direction of the total angular momentum \cite{PhysRevD.49.6274}. Spin-precession affects the gravitational waveform in three ways (1) it contributes to the orbital decay of the binary, and thus to the accumulated phase of the gravitational wave; (2) it causes the orbital plane to precess, changing its orientation relative to us and thus modulating the waveform; and finally (3) spin contributes directly to the gravitational wave amplitude through higher order terms in the post-Newtonian expansion\cite{PhysRevD.52.821}.

%\ros{not clear exactly what you are distinguishing here: is this the true current-multipole source term from the tranverse spin components, or the amplitude effect of orbital plane misalignment? More precise terminology, and citations, help. For the second, I'd cite Apostolatos et al 1994; for the first, either that or Kidder; for the 3rd, depends on what you mean}.

The identification and analysis of precessing systems can have a significant impact in our understanding of astrophysical binary formation channels. There are two primary ways that these binaries can form. The first is in the heart of star clusters \cite{Sigurdsson_1995, Antonini_2020, 10.1093/mnras/sty2211}: in these dense environments dynamic interactions between systems can lead to spins becoming misaligned; e.g. in young star clusters \cite{Trani_2021}. The second primary formation channel occurs in isolated systems in the galactic field \cite{Bethe_1998, Tanikawa_2021, Belczynski_2020}, where there are several possible mechanisms of formation. Of particular interest are binaries formed through supernova kicks, where it is estimated that as high as 80\% of objects can have misalignment angles greater than 30\degree \cite{Kalogera_2000}.

In order to rapidly identify spin-precession effects in gravitational wave signals, the raw output of parameter estimators such as RIFT is recast into single effective parameters such as $\chi_p$ (see Eq.(\ref{chip}) that judge the degree to which a binary is precessing. $\chi_p$ is a parameter that expresses the maximum misaligned spin in the binary, taking information from only one of the two objects. In this work we shall discuss the implementation in RIFT of the updated precession parameter  $\langle \chi_p \rangle$, first introduced in \cite{Gerosa_2021}. This new parameter is a more faithful characterization of a binary's precession, as it includes misalignment information from both objects in the binary and averages over all the angular variations on the precession timescale. It will be shown that there are significant differences in the valuation of precession between these two parameters, particularly for systems with large spin magnitudes $(\chi \geq 0.5)$ and even mass ratio $(m_1 \approx m_2)$. The expression of these differences will then be highlighted for select events from O3b, of which ten show large amounts of precession $\left(\langle \chi_p \rangle > 0.5\right)$.

%\ros{disambiguate from paper where it was introduced - shown there. Emphasize the novel discussions like Fig 2, 3,Section 4}

This work will be organized as follows. In Sec.(\ref{parameterization}) we review the parameterization of precession magnitude as a way to characterize the binary. Following \cite{Gerosa_2021}, in Sec.(\ref{implementation}) we describe the implementation of $\langle \chi_p \rangle$ in the RIFT parameter estimation algorithm. In Sec.(\ref{paramcompare}) the differences between the current standard parameter $\chi_p$ and the updated parameter $\langle \chi_p \rangle$ are discussed as functions of the intrinsic parameter space. Finally in Sec.(\ref{O3bresults}) the precession characteristics for all 36 events from the second half of the Advanced LIGO and Advanced Virgo third operating run (O3b) are reported and discussed.

\footnotetext{In LSC programming language this angle is often referred to as \texttt{phi12}.}

%------------------------------------------------

\section{Parameterizing Precession \label{parameterization}}

\begin{figure}[H]
    \centering
    \includegraphics[width=6cm]{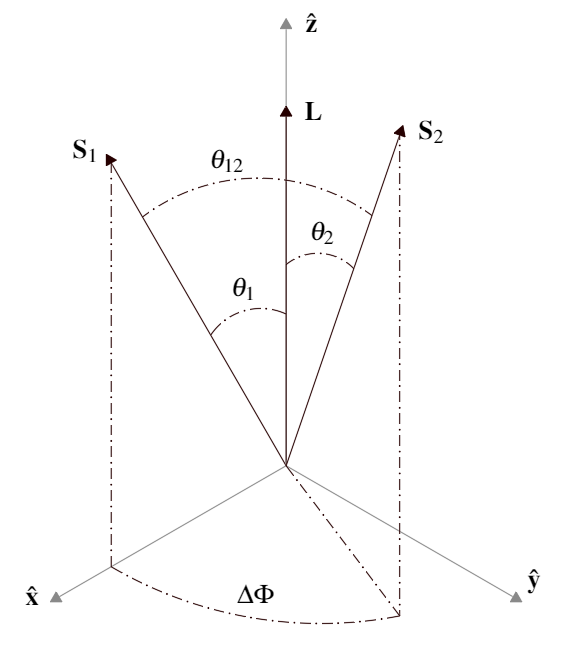}
    \caption[Caption for USC]{Coordinate system in the binary source frame. The direction of the angular momentum $\mathbf{L}$ is taken to be along the $\hat{z}$ axis, such that the orbital motion is confined to the x-y plane. The polar angles $\left(\theta_1, \theta_2\right)$ described the misalignment of the two object's spins $\left(\mathbf{S}_1, \mathbf{S}_2\right)$ with respect to the direction of orbital angular momentum. $\theta_{1 2}$ is the direct angle between these two spin directions, and $\Delta \Phi$ is the azimuthal difference between the two spin projections onto the orbital plane\protect\footnotemark. Note that $\mathbf{S_1}$ is defined to be constrained within the x-z plane; as such both polar angles range from $0 \leq \left(\theta_1, \theta_2\right) \leq \pi$ and the azimuthal angle difference ranges from $0 \leq \Delta \Phi \leq 2\pi$.}
    \label{UCS}
\end{figure}

In a binary system containing objects with arbitrarily oriented spins, the spin-precession behavior may be entirely defined by eight intrinsic parameters: the two object masses and six spin components $\left(m_1, m_2, \mathbf{S}_{1 x}, \mathbf{S}_{1 y}, \mathbf{S}_{1 z}, \mathbf{S}_{2 x}, \mathbf{S}_{2 y}, \mathbf{S}_{2 z}\right)$. To analyze such systems we take the coordinate system as shown in Fig.(\ref{UCS}), following \cite{Farr}. We also take the convention for the mass ratio $q \equiv m_2 / m_1 \leq 1, m_1 > m_2$.

%\LC{do not rely on the exact placement of figure 1 - rather point to figure 1 explicitely, and maybe move here some of what is in the caption }
%\ros{How is q defined: $m_2/m_1$? is $m_1>m_2$?}

In a coalescing binary the evolution of the orbital angular momentum per orbit proceeds as follows:
\begin{align}
    \dv{\mathbf{L}}{t} = \dv{\hat{\mathbf{L}}}{t} L + \dv{L}{t}\hat{\mathbf{L}} \label{eq1}
\end{align}
where the first term describes the \emph{orbital plane precession} - the change in the direction of the orbital angular momentum over time, and the second term describes the radiation reaction. These two effects happen on different timescales and can thus be decoupled from each other. The first scalar parameter introduced to characterize precession, presented in \cite{Schmidt_2015}, defines $\chi_p$ as:
\begin{align}
    \chi_p \equiv \max \left(\chi_1 \sin{\theta_1}, \tilde{\Omega} \chi_2 \sin{\theta_2}\right) \label{chip}
\end{align}
where $\tilde{\Omega}$ is the ratio of spin frequencies to leading order:

\begin{align}
    \tilde{\Omega} &= \frac{\Omega_2}{\Omega_1} = q \frac{4 q + 3}{4 + 3 q} + \mathcal{O}\left(\frac{M^2}{L}\right)\label{norm}
\end{align}

As defined in Eq.(\ref{chip}),
$\chi_p$ is the precession parameter has previously been widely reported by standard parameter estimation codes to assess the degree to which a binary is precessing. The normalization factor of $\tilde{\Omega}$ constrains this parameter to the region $0 \leq \chi_p \leq 1$, with a larger value indicating a larger degree of spin misalignment and thus orbital plane precession. This parameter is derived from the normalized magnitude of orbital plane precession, which we refer to as the generalized $\chi_p$:
%\ros{have you explained what this equation means? Where does it come from, what is it geometrically?}

\begin{align}
    \chi_{p, \text{gen.}} = \frac{1}{\Omega_1}\abs{\dv{\hat{\mathbf{L}}}{t}} =  &\left[\left(\chi_1 \sin{\theta_1}\right)^2 + \left(\tilde{\Omega}\chi_2 \sin{\theta_2}\right)^2 +\right.\nonumber\\
    &\left. 2 \tilde{\Omega} \chi_1 \chi_2 \sin{\theta_1} \sin{\theta_2} \cos{\Delta \Phi}\right]^{\frac{1}{2}} \label{gen}
\end{align}

by taking the average value of the $\cos{\Delta\Phi}$ extrema.  However as pointed out by Gerosa et al. \cite{Gerosa_2021} this parameter has several drawbacks. Consider that $\chi_p$ as defined in Eq.(\ref{chip}) effectively judges only the larger of the two spin projections onto the orbital plane; i.e. it takes information from only one of the two objects in the binary. Furthermore the reductive assumption of taking only the extrema of $\Delta \Phi$ is tantamount to averaging over only this angle; however the other two relevant angles $\theta_1, \theta_2$ vary on the same timescale. As all three angles vary on the same timescale $t_{pre} \propto \left(r/M\right)^{5/2}$, information is lost by averaging over only one of the three. A resolution to this problem involves averaging over all the angular variations on the precession timescale. This new parameterization takes the general definition of Eq.(\ref{gen}) and computes the average:
\begin{align}
    \langle \chi_p \rangle &= \frac{\int \chi_p\left(\psi\right) \left(\dv{\psi}{t}\right)^{-1} d\psi}{\int \left(\dv{\psi}{t}\right)^{-1} d\psi} \label{chipavg}
\end{align}
where $\psi(t)$ is chosen as a quantity that characterizes the one-parameter spin precession dynamics on the spin precession timescale. The averaged precession parameter $\langle \chi_p \rangle $ retains information from both objects in the binary, and is a constant of motion at 2PN. It is also constrained to the domain $0 \leq \langle \chi_p \rangle \leq 2$, with the region $1 < \langle \chi_p \rangle \leq 2$ exclusive to binaries with two misaligned spins. For the implementation in RIFT presented below, the total spin magnitude $S = \abs{\mathbf{S}_1 + \mathbf{S}_2}$ was chosen for $\psi$, as it is easily obtained from the intrinsic parameters without need for further transformation.

\section{Implementation \label{implementation}}

\subsection{RIFT  \label{riftreview}}

%\LC{spell out what RIFT means, the first time the acronym is used}

Rapid Iterative FiTting (RIFT) \cite{lange2018rapid, Rift_algorithm} is a parameter estimation algorithm that compares a candidate coalescing binary gravitational wave signal to existing waveforms, then marginalizes the likelihood of the signal data over the characteristic coordinates of the binary's coalescence event relative to the Earth. The basic procedure for using RIFT\footnotemark is simple. Metadata from a gravitational wave candidate event is taken directly from GraceDB (the Gravitational-Wave Candidate Event Database). The user then chooses an approximant; the type of simulation basis template against which the real data will be compared. RIFT also allows for several options, such as specifying the power spectral density (PSD), allowing for higher-order modes, number of iterations, etc. 

\footnotetext{For more information on using RIFT and associated functions, see \url{https://github.com/oshaughn/RIFT_tutorials}.}

% \LC{I think there is too much technical detail on how RIFT is run, too many script names}

From this user input, RIFT goes through two primary stages. The first is Integrate Likelihood Extrinsic (ILE), which evaluates the marginalized likelihood on candidate points using Monte Carlo integration. The second stage is Construct Intrinsic Posterior (CIP), which then estimates the likelihood and posterior distribution using a Gaussian process. This posterior is then used as a prior in the next iteration, and the two tasks repeat. The output of RIFT is a set of posterior probability distributions for the binary's eight intrinsic parameters and if specified, also the extrinsic parameters. For any specific sample within this distribution, one may then conduct any transformation dependent upon these parameters. For a trivial example one might compute the mass ratio $q = m_2/m_1$, and do so for every sample to create a posterior distribution.

%\LC{I am a little worried about using all these routine names (here and in the following) - those could change, especially if the code is not frozen. Can we just explain the process/math? Also some of the process could be moved to an appendix if needed}

The RIFT package provides many standard parameter transformations (\texttt{lalsimutils.py}), as well as a plotting tool (\texttt{plot\_posterior\_corner.py}) for creating corner plots of the posterior distributions.   We modified  these two utilities  to calculate and illustrate the averaged precession parameter $\langle \chi_p \rangle$.  We emphasize these calculations are purely postprocessing, performed after the main parameter estimation process of RIFT. As they require only  posterior samples, such calculations can be subsequently performed with any existing posterior data. These low-cost postprocessing transformations can be performed readily with local CPUs.

\subsection{Calculation of $\langle \chi_p \rangle$ in RIFT \label{calculation}}

To calculate $\langle \chi_p \rangle$ for a given set of eight intrinsic parameters, the following algorithm has been programmed into RIFT's \texttt{lalsimutils.py} analysis tool as an option in the \texttt{extract\_param} function. This closely follows the example implementation presented in \cite{Gerosa_2021} (and implemented with the
\texttt{precession} package \cite{Gerosa_2016}), with adjustments made to cooperate with RIFT's existing functions. It should be noted that the RIFT output samples for the two masses are given in the units of solar mass $M_\odot$; for the purposes of this calculation we therefore first convert these to units of $\left[\frac{s}{kg}\right]$ using the conversion factor $M_\odot \frac{G}{c^3}$ to reduce the numerical order across the rest of the algorithm. We begin by calculating the angles $\left(\theta_1, \theta_2, \Delta \Phi\right)$ using the built-in function \texttt{extract\_system\_frame}. This function first fixes the direction of orbital angular momentum (see Fig.(1)) and then computes the following:

\begin{align}
\theta_1 &= \arccos{\left(\hat{\mathbf{S}}_1 \cdot \hat{\mathbf{L}}\right)}\nonumber\\
\theta_2 &= \arccos{\left(\hat{\mathbf{S}}_2 \cdot \hat{\mathbf{L}}\right)}\nonumber\\
\Delta \Phi &= \Re{-i \ln{\left(\frac{\left(\hat{\mathbf{x}} + i \hat{\mathbf{y}}\right)\cdot\hat{\mathbf{S}}_1}{\left(\hat{\mathbf{x}} + i \hat{\mathbf{y}}\right)\cdot\hat{\mathbf{S}}_2}\right)}}
\end{align}

The function \texttt{extract\_system\_frame} also gives us the dimensionless spin magnitudes $\chi_i = \frac{S_i}{m_i^2}$. The mass ratio is defined as $q = \frac{m_2}{m_1}$ with $m_1 > m_2$. The last value needed is the reference frequency $f_{ref}$, which is given a default value of $20 [Hz]$, although this may be changed through user specification. The (Newtonian) orbital angular velocity is then $\omega = \pi f_{ref}$. We can now calculate the separation distance as per Eq.(4.13) in Ref. \cite{PhysRevD.52.821}:

\begin{align}
    \frac{r}{M} = &\left(M \omega\right)^{-\frac{2}{3}} - \left[1 - \frac{q}{3(1 + q)^2}\right]\nonumber\\
    &- \frac{(M\omega)^{\frac{1}{3}}}{3(1+q)^2} \left[(3q + 2)\chi_1 \cos{\theta_1} + q(3 + 2q)\chi_2 \cos{\theta_2}\right]\nonumber\\
    &+ \left(M \omega\right)^{\frac{2}{3}} \bigg[\frac{q}{(1+q)^2}\left(\frac{19}{4} + \frac{q}{9 (1+q)^2}\right)\nonumber\\
    &- \frac{\chi_1 \chi_2}{2}  \left(\sin{\theta_1}\sin{\theta_2}\cos{\Delta \Phi} - 2\cos{\theta_1}\cos{\theta_2}\right)\bigg]
\end{align}

We then proceed to calculate the magnitude of the angular momentum $L$, the magnitude of the total angular momentum $J$, and the dimensionless effective aligned spin $\xi$ \cite{Racine_2008, Damour_2001}:

\begin{align}
    L = m_1 m_2 \sqrt{\frac{r}{M}}
\end{align}

\begin{align}
    J = &\Big[L^2 + S_1^2 + S_2^2 + 2L\left(S_1 \cos{\theta_1} + S_2\cos{\theta_2}\right)\nonumber\\
    &+ 2 S_1 S_2 \left(\sin{\theta_1}\sin{\theta_2}\cos{\Delta \Phi} + \cos{\theta_1}\cos{\theta_2}\right)\Big]^{\frac{1}{2}}
\end{align}

\begin{align}
    \xi &\equiv M^{-2} \left[(1+q) \mathbf{S}_1 + (1 + q^{-1})\mathbf{S}_2\right]\cdot\hat{\mathbf{L}}\nonumber\\
    &= \frac{1+q}{q M^2} \left(q S_1 \cos{\theta_1} + S_2 \cos{\theta_2}\right)
\end{align}

Note that $\xi$ is a conserved quantity on both the precession and radiation-reaction timescales \cite{Gerosa_2016}. Recall that our goal is to calculate Eq.(\ref{chipavg}) using the total spin; this equation now becomes:

\begin{align}
    \langle \chi_p \rangle &= \frac{\int_{S_-}^{S_+} \chi_p\left(S\right) \left(\dv{S}{t}\right)^{-1} dS}{\int_{S_-}^{S_+} \left(\dv{S}{t}\right)^{-1} dS} \label{eq10}
\end{align}

The time derivative of $S$ is given by \cite{Racine_2008}:

\begin{align}
    \abs{\dv{S}{t}} = &\frac{3}{2} \frac{S_1 S_2 M^9}{L^5} \frac{q^5(1-q)}{(1+q)^11} \bigg[1 - \nonumber\\
     &\frac{q M^2 \xi}{L (1+q)^2}\bigg] \frac{\sin{\theta_1}(S) \sin{\theta_2}(S)\abs{\sin{\Delta \Phi(S)}}}{S}
\end{align}

Notice that this derivative contains angular parametric equations of $S$, which are given by:

\begin{align}
    \theta_1(S) &= \cos^{-1}{\left\{\frac{1}{2(1-q)S_1} \left[\frac{J^2 - L^2 - S^2}{L} - \frac{2 q M^2 \xi}{1 + q}\right]\right\}}\nonumber\\
    \theta_2(S) &= \cos^{-1}{\left\{\frac{1}{2(1-q)S_2} \left[- \frac{J^2 - L^2 - S^2}{L} - \frac{2 M^2 \xi}{1 + q}\right]\right\}}\nonumber\\
    \Delta \Phi(S) &=  \cos^{-1}{\left\{\frac{S^2 - S_1^2 - S_2^2 - 2 S_1 S_2 \cos{\theta_1}(S) \cos{\theta_2}(S)}{2 S_1 S_2 \sin{\theta_1}(S) \sin{\theta_2}(S)}\right\}}
\end{align}

The limits of integration $S_{\pm}$ in Eq.(15) correspond to the two extremal solutions of $\dv{S}{t} = 0$. To obtain these we use the function \texttt{Sb\_limits}, which is part of the \texttt{precession} package \cite{Gerosa_2016}. This evaluates the geometrical constraints of the system based on the following definitions. As shown in Fig.(\ref{UCS}), the two polar angles range from $0 \leq \left(\theta_1, \theta_2\right) \leq \pi$ and the azimuthal angle difference ranges from $0 \leq \Delta \Phi \leq 2\pi$. Based on the limits of the polar angles and azimuthal angle difference, the effective spin $\xi$, along with the total spin $S$, and the total angular momentum $J$, also have geometric limits:

\begin{align}
    &- \frac{1+q}{M^2}\left(S_1 + \frac{S_2}{q}\right) \leq \xi \leq \frac{1+q}{M^2} \left(S_1 + \frac{S_2}{q}\right)\nonumber\\
    &\abs{S_1 - S_2} \leq S \leq S_1 + S_2\nonumber\\
    &\max \left(0, L - S_1 - S_2, \abs{S_1 - S_2} - L\right) \leq J \leq L + S_1 + S_2
\end{align}

Notice that these constraints are dependent on each other. To solve for the limits $S_{\pm}$, the effective potentials $\xi_{\pm}$ \cite{Kesden_2015} for spin precession are used:

\begin{align}
    \xi_{\pm} &= \frac{1}{4 q M^2 S^2 L} \bigg\{\left(J^2 - L^2 - S^2\right)\big[S^2 (1+q)^2 - (S_1^2\nonumber\\
    &- S_2^2)(1-q^2)\big] \pm (1-q^2)\Big(\left[J^2 - (L-S)^2\right]\big[(L\nonumber\\
    &+S)^2 - J^2\big]\left[S^2 - (S_1 - S_2)^2\right]\left[(S_1+S_2)^2 - S^2\right]\Big)^{\frac{1}{2}} \bigg\}
\end{align}

The solutions $S_{\pm}$ to the equations $\xi_{\pm} - \xi = 0$ are then found using \texttt{scipy.optimize.brentq}, an implementation of Brent's method for root finding. Each of the integrals in Eq.(15) may now be calculated, and are computed by \texttt{scipy.integrate.quad}. Note that the precession cycle $S_- \rightarrow S_+ \rightarrow S_-$ is symmetric; as such we need only integrate over half of the total cycle (i.e. from $S_-$ to $S_+$).

Given the input of the eight intrinsic binary parameters this algorithm computes a value for $\langle \chi_p \rangle$ which lies in the interval $[0,2]$. However it should be noted that this procedure does fail for a few specific cases. If both spins are entirely aligned (or anti-aligned) with the orbital angular momentum, then the above algorithm will fail. However this is a trivial case, as if there is no misalignment then there is no precession and we set $\langle \chi_p \rangle = 0$. If one of the spins is exactly zero (i.e. $S_i = 0$) but the other spin is not, then this algorithm also fails. In this case $\langle \chi_p \rangle = \chi_p$, and this is the parameter that is calculated. This equivalence can be seen by applying the single misaligned spin condition to Eq.(\ref{gen}) for the generalized precession parameter. Here taking $\chi_1 \neq 0, \chi_2 = 0$:

\begin{align}
    \chi_{p, \text{gen.}} &= \left[\left(\chi_1 \sin{\theta_1}\right)^2\right]^{\frac{1}{2}}\nonumber\\
    &= \chi_1 \sin{\theta_1}
\end{align}

Or the opposite case, where $\chi_1 = 0, \chi_2 \neq 0$:

\begin{align}
    \chi_{p, \text{gen.}} &= \left[\left(\tilde{\Omega}\chi_2 \sin{\theta_2}\right)^2\right]^{\frac{1}{2}}\nonumber\\
    &= \tilde{\Omega}\chi_2 \sin{\theta_2}
\end{align}

Hence a comparison to Eq.(\ref{chip}) shows that for either case of single misaligned spin calculation of the general precession parameter reduces to simply a calculation of $\chi_p$. This is true also for the averaged precession parameter $\langle \chi_p \rangle$; as the magnitude of one of the spins approaches zero, $\langle \chi_p \rangle$ is well approximated by $\chi_p$.

RIFT's supplemental tool for plotting posterior probability distributions is \texttt{plot\_posterior\_corner.py}, which reads in the posterior samples produces by the ILE/CIP process and creates corner plots of the total probability distribution for the binary. Within this script one can call any of the transformations programmed into \texttt{lalsimutils.py} to similarly create posteriors for parameters such as $\langle \chi_p \rangle$. This parameter has been included in the latest release of RIFT, and posteriors for both precession parameters can now be created for any existing and future gravitational wave event data.

\section{Parameter Comparison \label{paramcompare}}

Let us now examine the key differences between these two precession parameters. The current standard parameter $\chi_p$ looks at the projection of the two object's spins onto the orbital plane, and takes the maximum of these two values. This produces a parameter value normalized to the domain $0 \leq \chi_p \leq 1$. However by taking this maximum we are considering information from only one of the objects in the binary, and using this information alone to judge spin misalignment and thus orbital plane precession. This method, while useful in specific cases, loses the $\cos{\Delta\phi}$ cross term from the generalized precession parameterization given in Eq.(\ref{gen}).

The new parameter $\langle \chi_p \rangle$ retains this term and averages over $\Delta \Phi$ on the precession timescale, and as such retains the spin-misalignment information from both objects in the binary. This produces a parameter value normalized to the domain $0 \leq \langle \chi_p \rangle \leq 2$. As pointed out in \cite{Gerosa_2021}, it is of crucial interest to note that any value of $\langle \chi_p \rangle > 1$ must necessarily correspond to a set of binary intrinsics where \emph{both} objects have misaligned spin. This is an immediately observable feature that is not present in $\chi_p$, as from this parameter there can be no direct correlation to the intrinsic space of both objects.

Based on the domains of these two parameters, at first glance one might expect that for a given set of source parameters $\langle \chi_p \rangle$ will yield values greater than $\chi_p$. However the inclusion of the cross term which contains $\cos{\Delta \Phi}$ in Eq.(\ref{gen}) means that for specific binary morphologies the averaged parameter can actually be much less than $\chi_p$. To highlight this, a set of $N=1000$ uniformly random binaries was generated with spin components $0 \leq \abs{S_i} \leq 1$. For each of these binaries both $\langle \chi_p \rangle$ and $\chi_p$ were calculated at fixed total mass, for mass ratios $q = \left\{1.00, 0.80, 0.50\right\}$ at a reference frequency of $f_{ref} = 20\: [\text{Hz}]$. It should be noted that the spin component values were randomized with a fixed seed; the only variable changing between the three tests below is the mass ratio. The data from these calculations are shown in Fig.(\ref{chip_v_chipavg}) below.

\begin{figure*}
    \centering
    \includegraphics[width=13cm]{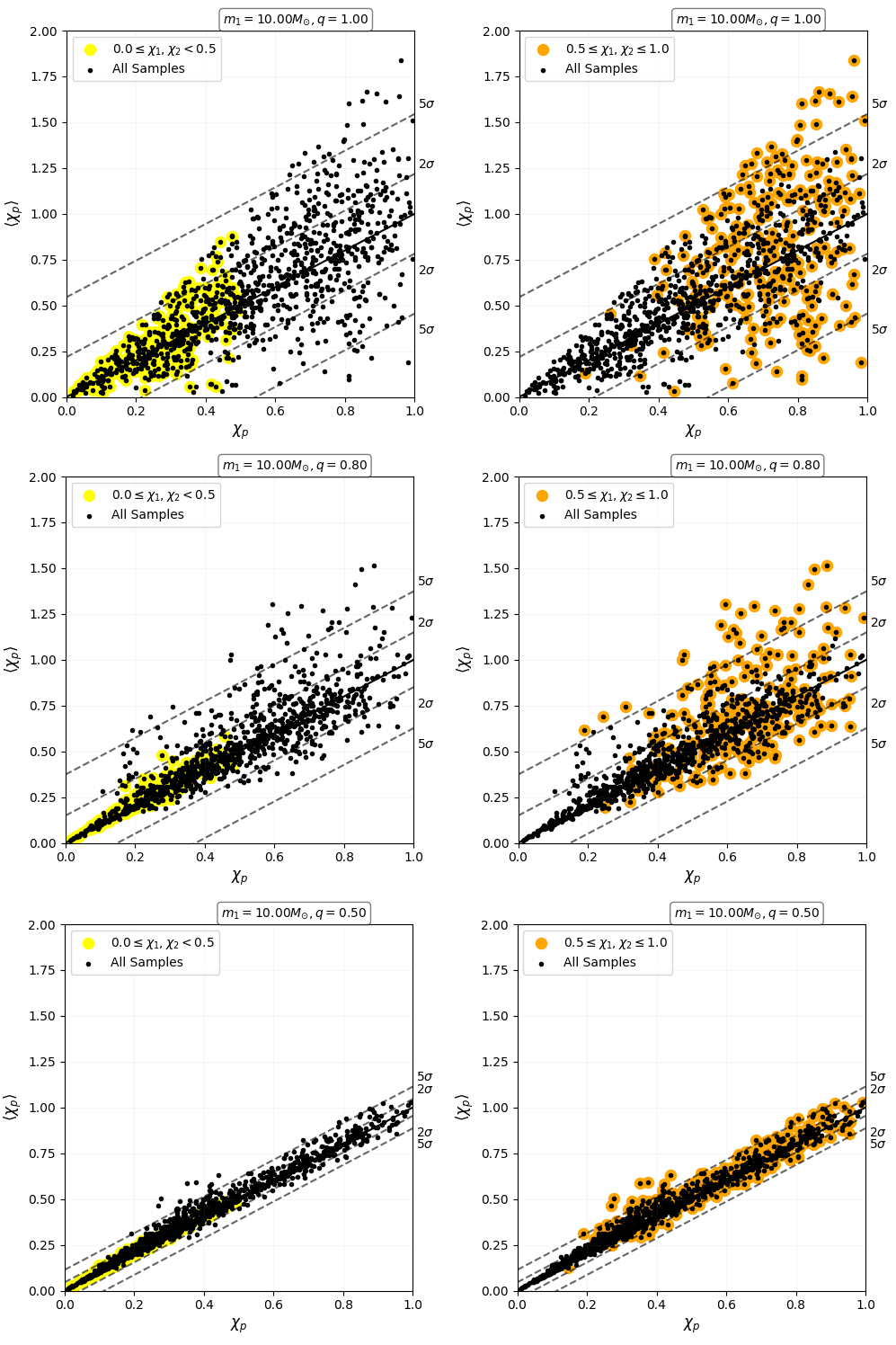}
    \caption{Precession parameter comparison for $N=1000$ uniformly random spin components $0 \leq \abs{S_i} \leq 1$ at $m_1 = 10.00\: \left[M_{\odot}\right]$ and $q = \left\{1.00, 0.80, 0.50\right\}$ with a reference frequency of $f_{ref} = 20\: [\text{Hz}]$. The yellow markers indicate samples with low spin ($0.0 \leq \chi_1, \chi_2 \leq 0.5$). The orange markers indicate samples with high spin ($0.5 \leq \chi_1, \chi_2 \leq 1.0$). The regions bounded by dashed lines indicate standard deviations $\sigma = \left\{0.109, 0.075, 0.023\right\}$ away from the solid black line $\chi_p = \langle \chi_p \rangle$ for each respective case.}
    \label{chip_v_chipavg}
\end{figure*}

\begin{figure*}
    \centering
    \includegraphics[width=\textwidth]{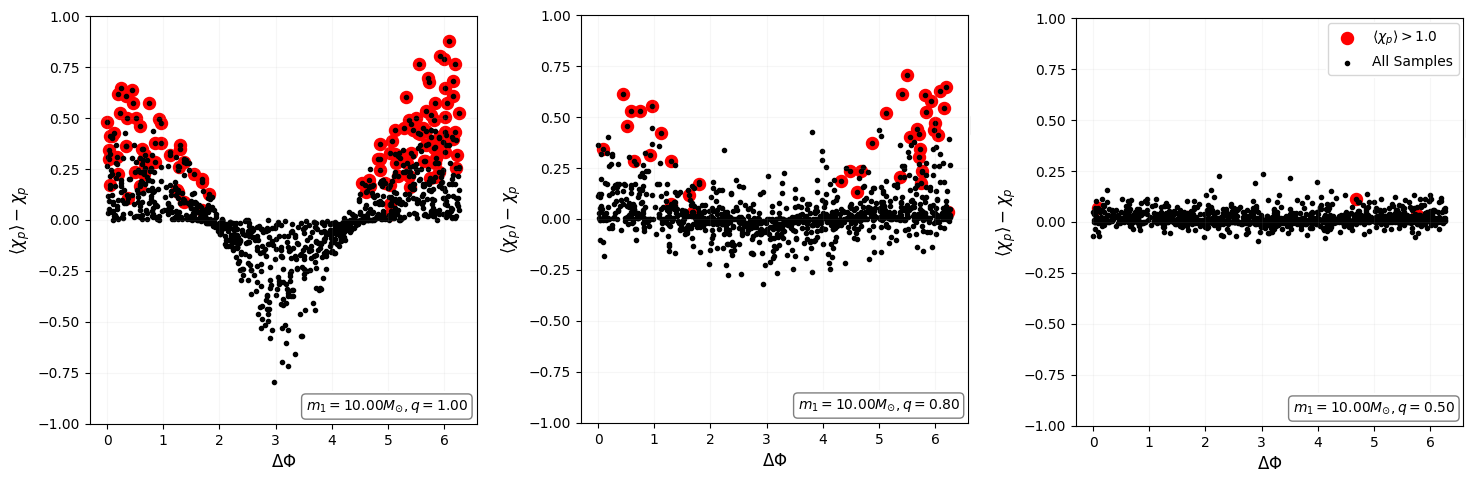}
    \caption{Precession parameter difference $\langle \chi_p \rangle - \chi_p$ vs. $\Delta \Phi$ for $N=1000$ uniformly random spin components $0 \leq \abs{S_i} \leq 1$ at $m_1 = 10.00 M_{\odot}$ and $q = \left\{1.00, 0.80, 0.50\right\}$ with a reference frequency of $f_{ref} = 20\: [\text{Hz}]$. The red markers indicate samples with $\langle \chi_p \rangle > 1$, corresponding to binaries which necessarily have two misaligned spins.}
    \label{diff_phi12}
\end{figure*}

One can see that for binaries in which both objects have relatively high spin $\left(0.5 \leq \chi_1, \chi_2 \leq 1.0\right)$ the distribution of the parameter space will be more widely spread about the $\chi_p = \langle \chi_p \rangle$ line, in some cases surpassing a $5 \sigma$ deviation. For binaries in which both objects have a relatively low spin  $\left(0.0 \leq \chi_1, \chi_2 \leq 0.5\right)$ the distribution of the parameter space will be less widely spread about the $\chi_p = \langle \chi_p \rangle$ line, never exceeding $5\sigma$. This difference relationship also depends strongly on the mass ratio $q$, but not on the total mass. The mass ratio enters as a variable in the normalizing factor $\tilde{\Omega}$ (Eq.(\ref{norm})). One can see that increasing the asymmetry in $q$ reduces the overall spread of the parameter difference for all spin values. At higher inverted mass ratios $(1/q \geq 2)$ the parameter difference is negligible, and in the limit as $q\to 0$ the two parameters converge. The parameter difference also depends strongly on the azimuthal angle difference $\Delta \Phi$. This variable enters into Eq.(\ref{gen}) as a cosine term, and this behavior is reflected in the distribution as shown in Fig.(\ref{diff_phi12}).

Notice that for $q = 1$ the distribution of values tends to follow the behavior of $\cos{\Delta\phi}$, but as the mass ratio decreases the distribution appears more uniform. Consider first the $q = 1$ image. The samples with the largest positive parameter difference cluster around the $\delta\phi = 0, 2\pi$ regions, corresponding to spin morphologies where the in-plane components of $\mathbf{S}_1, \mathbf{S}_2$ are aligned. These regions, where $\cos{\Delta\phi} > 0$ also represent the domain in which $\langle \chi_p \rangle > 1$. These samples, highlighted in red, correspond to binaries that necessarily have two misaligned spins - qualitative information that is not conveyed by $\chi_p$. The samples with the largest negative difference cluster in the region surrounding $\Delta \phi = \pi$, where $\cos{\Delta\phi} < 0$. This region corresponds to spin morphologies where the in-plane components of $\mathbf{S}_1, \mathbf{S}_2$ are anti-aligned. The orthogonal conditions for in-plane spin components $\left(\Delta\phi = \pi/2, 3\pi/2\right)$ correspond to roots in $\langle \chi_p \rangle - \chi_p$ vs. $\Delta \phi$, indicating that for these morphologies the two parameters are equivalent. However these behaviors degrade as the mass ratio decreases. Progressing from left to right in Fig.(\ref{diff_phi12}), the cosine-like behavior of $\langle \chi_p \rangle - \chi_p$ vs. $\Delta \Phi$ loses definition and approaches a uniform distribution. Additionally the quantity of samples with $\langle \chi_p \rangle > 1$ is significantly reduced, indicating that this region is less useful as a qualitative heuristic for double-misalignment as the mass ratio deviates from unity.

%\ros{Put this simple derivation earlier, perhaps in section 2}
These results demonstrate that for practically any given spin morphology, the parameter $\chi_p$ either under or overestimates the binary's precession. This is because $\chi_p$ inherently takes information from only one object in the binary, and does not take into account all the angular variations of the two objects over the precession timescale. However, $\chi_p$ is still a useful parameter when the mass asymmetry is large, or if one of the two spins is close to zero where $\langle \chi_p \rangle = \chi_p$ as discussed in Sec.(\ref{calculation}).

\section{Precession Results from GWTC-3 \label{O3bresults}}

Presented below are the precession parameter posteriors for all 36 events from the second half of the Advanced LIGO and Advanced Virgo third observing run, as presented in GWTC-3 \cite{gwtc3}. The intrinsic parameter data used to calculate these come from the public release \cite{gwtc3_data, gwopendata}, and were converted to RIFT readable format using the PESummary python package \cite{Hoy2020vys}. From this release the \texttt{SEOBNRv4PHM} \cite{SEOBNRv4PHM} approximant samples created by RIFT (prior to spin evolution and cosmological re-weighting) were analyzed. These data were chosen to be as close to the raw RIFT output as possible, differing only by standard calibration adjustment. As such the data shown here may differ from those reported in the O3b catalog, which use a combination of samples from both the time-domain parameter estimation (RIFT, \texttt{SEOBNRv4PHM}) and frequency-domain parameter estimation (Bilby, \texttt{IMRPhenomXPHM} \cite{IMRPhenomXPHM}). Presented here are posterior probability distributions for both $\chi_p$ and $\langle \chi_p \rangle$, displayed in Fig.(\ref{O3bEvents}) below, with median and 90\% confidence intervals reported in Table(\ref{O3b_table}).

\begin{figure*}
    \centering
    \includegraphics[width=16cm]{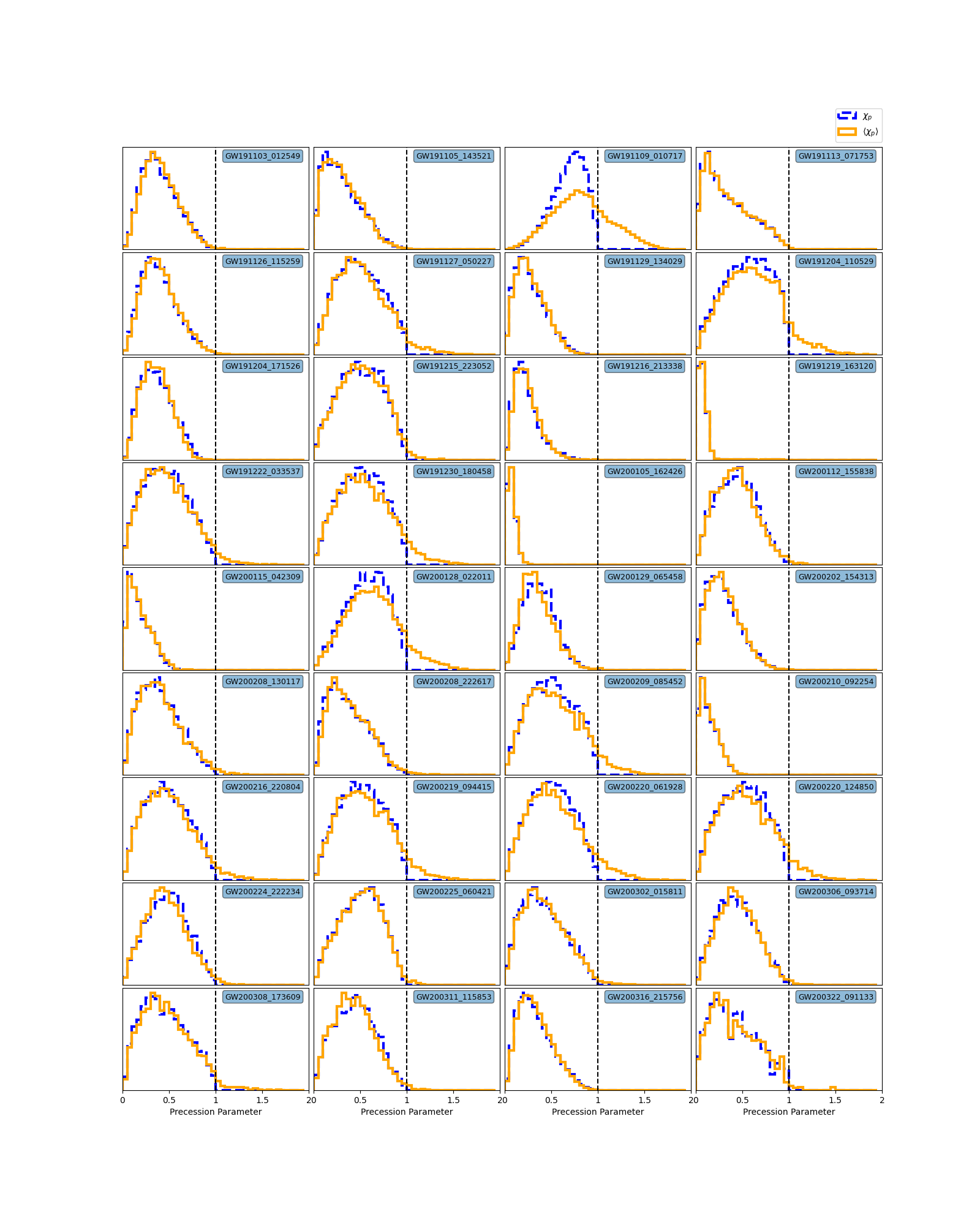}
    \caption{Posterior probability distributions for $\chi_p$ and $\langle \chi_p \rangle$ for all 36 O3b events. The dashed line marks $\langle \chi_p \rangle > 1$; this region is exclusive to binaries with two misaligned spins.}
    \label{O3bEvents}
\end{figure*}

Note that for the majority of these events there are not significant differences in the posterior distributions, as there is not a significant amount of precession. However there are a number of events that have a large probability of being highly precessing, with a parameter peak $\geq 0.5$. The posterior probability distributions for these events show a significant tail in the $\langle \chi_p \rangle > 1$ region, corresponding to the probability that the originating system necessarily contains two misaligned spins.These highly precessing events are show in Fig.(\ref{highprecessionevents}) with $P\left(\langle \chi_p \rangle > 1\right)$ listed in Table(\ref{highp_table}) below.  

%\LC{where do I see this event? it is lost in the many plots of figure 3 - could show a subset of posterior for the highly processing event as is done for the table}\\

\renewcommand{\arraystretch}{1.5}
\begin{table}[H]
\caption{Highly Precessing Events from O3b}
\begin{tabular}{|c|c|c|c|ll}
\cline{1-4}
\textbf{Event} & $\mathbf{\chi_p}$ & $\mathbf{\langle \chi_p \rangle}$ & \multicolumn{1}{l|}{ $\mathbf{P\left(\langle \chi_p \rangle > 1\right)}$}   \\ \cline{1-4}
GW191109\_010717      &   $0.71^{+0.23}_{-0.36}$   & $0.81^{+0.57}_{-0.48}$  & 0.28  \\ \cline{1-4}
GW191204\_110529      & $0.57^{+0.35}_{-0.42}$     &  $0.60^{+0.53}_{-0.44}$   & 0.09        \\ \cline{1-4}
GW191215\_223052      &  $0.50^{+0.37}_{-0.38}$    &    $0.51^{+0.39}_{-0.39}$    & 0.01                       \\ \cline{1-4}
GW191230\_180458      &  $0.52^{+0.38}_{-0.39}$    &    $0.53^{+0.52}_{-0.40}$    & 0.07                       \\ \cline{1-4}
GW200128\_022011      &  $0.58^{+0.32}_{-0.39}$    &  $0.62^{+0.52}_{-0.43}$     & 0.10                        \\ \cline{1-4}
GW200209\_085452      &  $0.50^{+0.40}_{-0.36}$    &  $0.51^{+0.54}_{-0.38}$     & 0.07                       \\ \cline{1-4}
GW200219\_094415      &  $0.52^{+0.38}_{-0.38}$    &  $0.53^{+0.51}_{-0.38}$   & 0.06                    \\ \cline{1-4}
GW200220\_061928      &  $0.50^{+0.37}_{-0.36}$    &  $0.51^{+0.56}_{-0.38}$   & 0.07                    \\ \cline{1-4}
GW200220\_124850      &  $0.52^{+0.38}_{-0.39}$    &  $0.52^{+0.53}_{-0.39}$   & 0.07                    \\ \cline{1-4}
GW200225\_060421      & $0.52^{+0.34}_{-0.38}$     &    $0.52^{+0.35}_{-0.38}$       & 0.01                    \\ \cline{1-4}
\end{tabular}
\label{highp_table}
\end{table}

Note that the information about dual-misalignment probability found from calculating $P\left(\langle \chi_p \rangle > 1\right)$ is not present from calculating $\chi_p$ alone. From these initial results we see that for highly precessing events this probability region offers valuable insight into the originating system, as evidence of dual-misaligned systems may contribute to constraining formation channels of precessing systems. Of particular interest is GW191109\_010717, shown below in Fig.(\ref{GW191109}) which presents with a sharply peaked $\chi_p = 0.71^{+0.23}_{-0.36}$, but a relatively flat distribution of $\langle \chi_p \rangle = 0.81^{+0.57}_{-0.48}$. There are a number of interesting features with this event that can affect the difference in distributions between $\chi_p$ and $\langle \chi_p \rangle$.\\

%the numbers in this paragraph have been updated 11-19 to reflect source frame
GW191109\_010717 shows component black holes that have dimensionless spin magnitudes of $\chi_1 = 0.83^{+ 0.15}_{-0.39}$, $\chi_2 = 0.57^{+0.39}_{-0.51}$. The mass ratio of the system is $q = 0.74^{+0.21}_{-0.25}$. As discussed in Sec(\ref{paramcompare}) above, one should expect significant differences between $\chi_p$ and $\langle \chi_p \rangle$ for a system with two large spin magnitudes and a mass ratio close to unity. Additionally both spins show moderately large probabilities of misalignment, with $\cos{\theta_1} = -0.44^{+0.61}_{-0.48}$, $\cos{\theta_2} = -0.32^{+0.99}_{-0.61}$. Recall that $\chi_p$ does not track the misalignment of both objects, only the maximum misalignment, while $\langle \chi_p \rangle$ preserves all the misalignment information. Furthermore, GW191109\_010717 is an interesting event due to the total mass. Its constituent black holes have masses of (in the source frame): $m_1 = 63^{+12}_{-10}\: \left[M_{\odot}\right]$, $m_2 = 46^{+11}_{-12}\: \left[M_{\odot}\right]$, leading to a final black hole of mass $M_f = 106^{+14}_{-14}\: \left[M_{\odot}\right]$. This places the final black hole in the intermediate mass black hole (IMBH) mass range $\left(10^2 - 10^5\: \left[M_{\odot}\right]\right)$.\\

% mass numbers updated 11-19 to reflect source frame
There are some remarkable similarities in the precession parameter distributions to those of the salient IMBH candidate event from the first half of the Advanced LIGO and Advanced Virgo third operating run (O3a) , GW190521 \cite{gwtc2}. Using parameter estimation data from the public release of GWTC-2 (also without cosmological reweighting), this event also shows a sharply peaked $\chi_p = 0.73^{+0.22}_{-0.41}$ with a $\langle \chi_p \rangle = 0.82^{+0.56}_{-0.49}$ that extends significantly into the $\langle \chi_p \rangle > 1$ region, yielding $P\left(\langle \chi_p \rangle > 1\right) = 0.25$. Similar to GW191109\_010717, GW190521 shows component black holes that have dimensionless spin magnitudes of $\chi_1 = 0.80^{+ 0.18}_{-0.58},$ $\chi_2 = 0.54^{+0.41}_{-0.48}$, and a mass ratio of the system is $q = 0.74^{+0.23}_{-0.42}$. Thus it also falls into the behavior pattern discussed in Sec(\ref{paramcompare}). Furthermore this event also shows high total mass, with constituent black holes of (in the source frame) $m_1 = 99^{+42}_{-19}\: \left[M_{\odot}\right]$ and $m_2 = 71^{+21}_{-28}\: \left[M_{\odot}\right]$, leading to a final remnant mass of $M_f = 162^{+35}_{-22}\: \left[M_{\odot}\right]$.\\

The high total mass of these two events leads to a very short signal duration (~0.1 s) in the LIGO and Virgo observable band. As pointed out in \cite{GW190521props}, the short signal duration of GW190521 limits our ability to judge the spin evolution of the system. As such the origin of the difference in distribution of precession parameters as shown in Figs.(\ref{GW191109}, \ref{GW190521}) may be the result of lack of information affecting our ability to characterize precession in IMBH systems. This topic will be addressed in a follow-up paper with further analysis of these two events.

%\ros{almost begging the question of what an injection would look like that's not precessing but equally short...or even zero spin.  Could they be this wide in this parameter?}

%notes on total masses

%GW191109_010717 : M = 144.33^{+16.17}_{-16.7}
%GW191204_171526 : M = 22.62^{+1.63}_{-0.38}
%GW191215_223052 : M = 58.67^{+5.6}_{-3.68}
%GW200128_022011 : M = 121.62^{+17.3}_{-15.53}
%GW200209_085452 : M = 101.47^{+20.92}_{-15.64}
%GW200219_094415 : M = 104.56^{+16.05}_{-12.86}
%GW200225_060421 : M = 41.49^{+3.23}_{-4.18}

\begin{figure*}
    \centering
    \includegraphics[width=\textwidth]{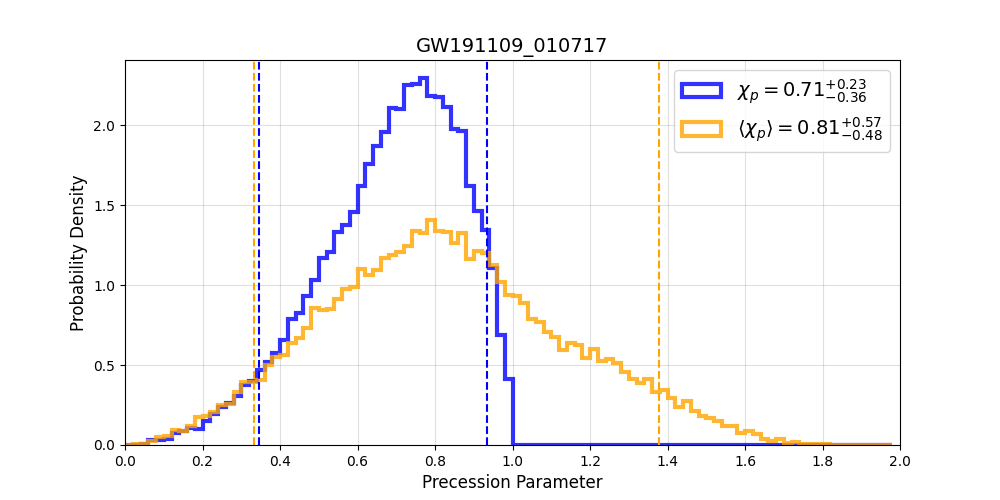}
    \caption{Precession parameter posterior probability distributions for GW191109\_010717, with median and 90\% confidence intervals.}
    \label{GW191109}
\end{figure*}

\begin{figure*}
    \centering
    \includegraphics[width=\textwidth]{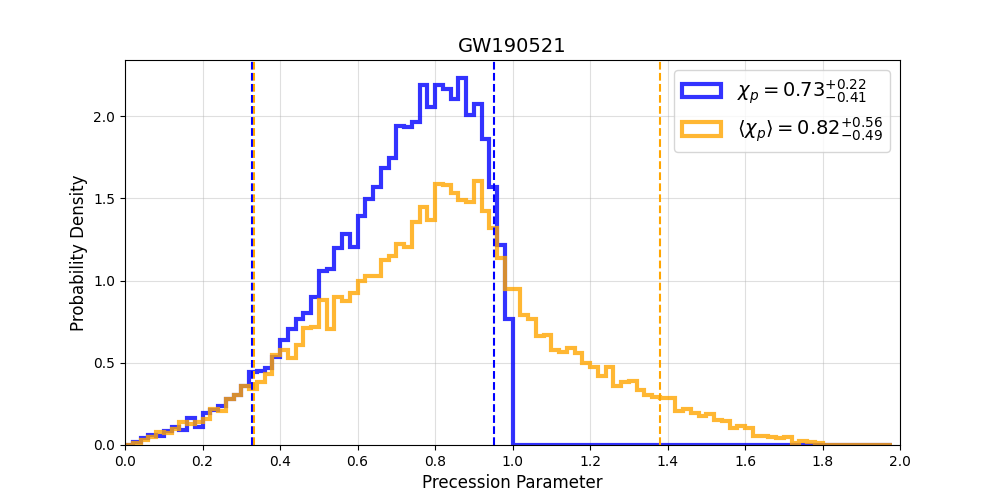}
    \caption{Precession parameter posterior probability distributions for GW190521, with median and 90\% confidence intervals.}
    \label{GW190521}
\end{figure*}

\begin{figure*}
    \centering
    \includegraphics[width=14cm]{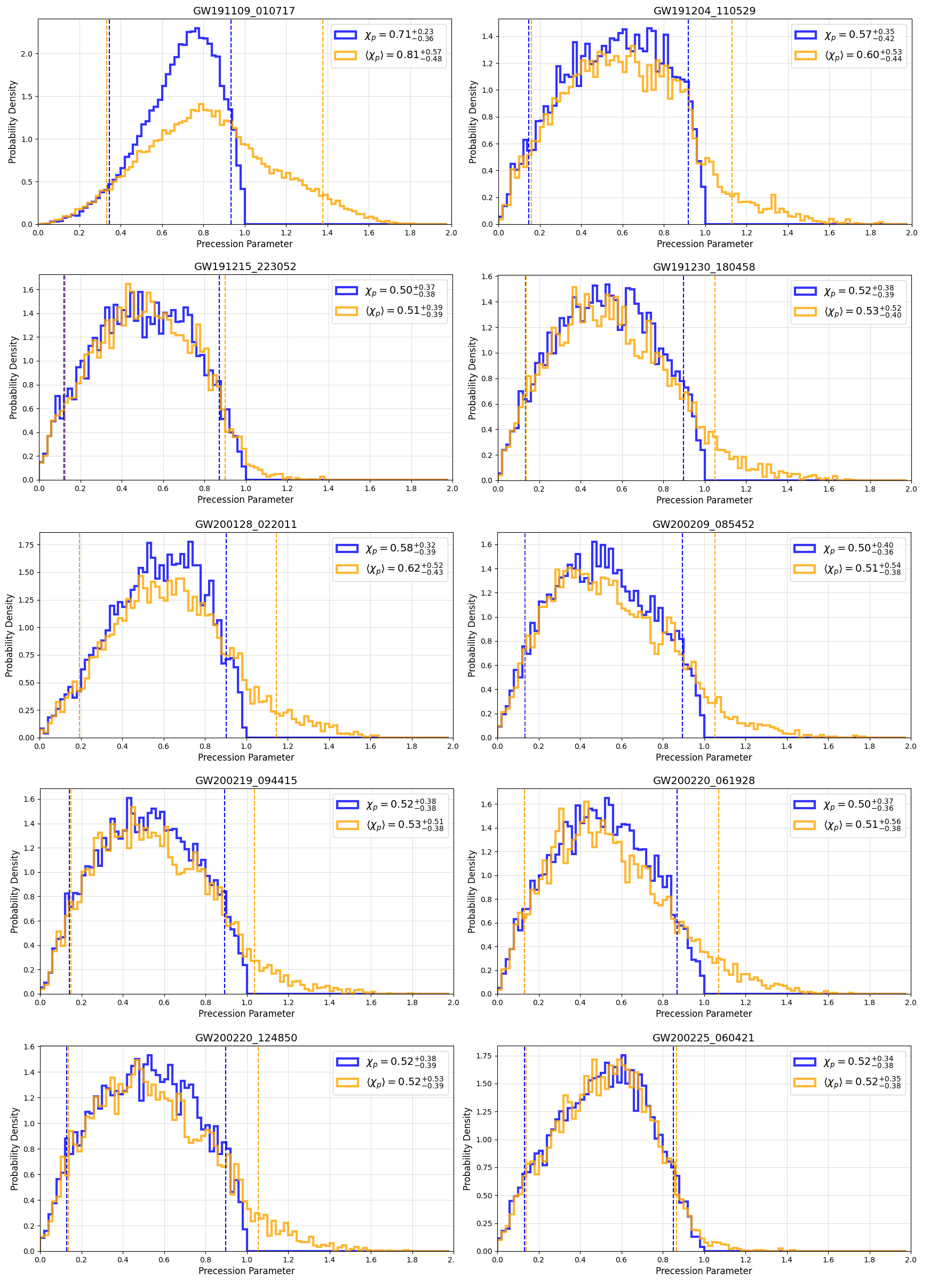}
    \caption{Precession parameter posterior probability distributions for the seven highly precessing ($\langle \chi_p \rangle > 0.5)$ events from O3b, with median and 90\% confidence intervals.}
    \label{highprecessionevents}
\end{figure*}

\renewcommand{\arraystretch}{1.5}
\begin{table*}
\centering
\caption{Precession Parameter Results for O3b Events}
\begin{tabular}{|l|l|l|||l|l|l|ll}
\cline{1-6}
\textbf{Event} & $\mathbf{\chi_p}$ & $\mathbf{\langle \chi_p \rangle}$ & \textbf{Event} & $\mathbf{\chi_p}$ & \multicolumn{1}{l|}{$\mathbf{\langle \chi_p \rangle}$}   \\ \cline{1-6}
GW191103\_012549      &   $0.38^{+0.38}_{-0.25}$   & $0.39^{+0.36}_{-0.25}$  & GW200129\_065458      &  $0.36^{+0.30}_{-0.24}$    &  $0.33^{+0.34}_{-0.22}$  \\ \cline{1-6}
GW191105\_143521      &   $0.28^{+0.41}_{-0.22}$   & $0.28^{+0.39}_{-0.23}$  & GW200202\_154313      &  $0.27^{+0.36}_{-0.21}$    &  $0.28^{+0.34}_{-0.22}$ \\ \cline{1-6} %up to here
GW191109\_010717      &   $0.71^{+0.23}_{-0.36}$   & $0.81^{+0.57}_{-0.48}$  & GW200208\_130117      &  $0.38^{+0.40}_{-0.28}$    &  $0.38^{+0.46}_{-0.28}$  \\ \cline{1-6}
GW191113\_071753      &  $0.29^{+0.51}_{-0.25}$    &  $0.30^{+0.51}_{-0.25}$   &  GW200208\_222617      &  $0.34^{+0.40}_{-0.26}$    &  $0.36^{+0.43}_{-0.26}$  \\ \cline{1-6}
GW191126\_115259      & $0.37^{+0.37}_{-0.25}$     &  $0.39^{+0.37}_{-0.25}$   &  GW200209\_085452      &  $0.50^{+0.40}_{-0.36}$    &  $0.51^{+0.54}_{-0.38}$   \\ \cline{1-6}
GW191127\_050227      & $0.48^{+0.41}_{-0.35}$     & $0.49^{+0.50}_{-0.35}$    &  GW200210\_092254      &  $0.12^{+0.21}_{-0.10}$    &  $0.13^{+0.21}_{-0.10}$  \\ \cline{1-6}
GW191129\_134029      & $0.25^{+0.34}_{-0.19}$     &  $0.26^{+0.31}_{-0.19}$    &  GW200216\_220804      &  $0.46^{+0.40}_{-0.34}$    &  $0.47^{+0.49}_{-0.34}$     \\ \cline{1-6}
GW191204\_110529      & $0.57^{+0.35}_{-0.42}$     &  $0.60^{+0.53}_{-0.44}$     & GW200219\_094415      &  $0.52^{+0.38}_{-0.38}$    &  $0.53^{+0.51}_{-0.38}$ \\ \cline{1-6}
GW191204\_171526      & $0.35^{+0.31}_{-0.23}$     &  $0.36^{+0.28}_{-0.23}$      & GW200220\_061928      &  $0.50^{+0.37}_{-0.36}$   &   $0.51^{+0.56}_{-0.38}$    \\ \cline{1-6}
GW191215\_223052      &  $0.50^{+0.37}_{-0.38}$    &    $0.51^{+0.39}_{-0.39}$     & GW200220\_124850      &  $0.52^{+0.38}_{-0.39}$    &  $0.52^{+0.53}_{-0.39}$  \\ \cline{1-6}
GW191216\_213338      &  $0.21^{+0.30}_{-0.14}$    &  $0.22^{+0.27}_{-0.14}$   &  GW200224\_222234      &  $0.48^{+0.35}_{-0.35}$    &    $0.45^{+0.41}_{-0.32}$  \\ \cline{1-6}
GW191219\_163120      &  $0.07^{+0.08}_{-0.05}$    &  $0.07^{+0.08}_{-0.05}$  &  GW200225\_060421      & $0.52^{+0.34}_{-0.38}$     &    $0.52^{+0.35}_{-0.38}$   \\ \cline{1-6}
GW191222\_033537      &  $0.45^{+0.40}_{-0.35}$    &  $0.45^{+0.48}_{-0.35}$       &   GW200302\_015811      &  $0.40^{+0.43}_{-0.30}$    &   $0.41^{+0.45}_{-0.31}$ \\ \cline{1-6}
GW191230\_180458      &  $0.52^{+0.38}_{-0.39}$    &  $0.53^{+0.52}_{-0.40}$      &  GW200306\_093714      &  $0.43^{+0.38}_{-0.31}$    &   $0.44^{+0.39}_{-0.31}$  \\ \cline{1-6}
GW200105\_162426      &  $0.07^{+0.09}_{-0.05}$    &  $0.07^{+0.09}_{-0.05}$     &  GW200308\_173609      &  $0.43^{+0.44}_{-0.33}$    &    $0.45^{+0.49}_{-0.34}$   \\ \cline{1-6}
GW200112\_155838      &  $0.44^{+0.36}_{-0.33}$    &  $0.42^{+0.39}_{-0.31}$        & GW200311\_115853      &  $0.45^{+0.36}_{-0.34}$    &   $0.43^{+0.39}_{-0.32}$   \\ \cline{1-6}
GW200115\_042309      &  $0.16^{+0.24}_{-0.12}$    &  $0.17^{+0.26}_{-0.13}$     & GW200316\_215756      &  $0.30^{+0.38}_{-0.21}$    &  $0.31^{+0.36}_{-0.22}$ \\ \cline{1-6}
GW200128\_022011      &  $0.58^{+0.32}_{-0.39}$    &  $0.62^{+0.52}_{-0.43}$    & GW200322\_091133      & $0.38^{+0.50}_{-0.31}$     &  $0.39^{+0.51}_{-0.31}$ \\ \cline{1-6} 
\end{tabular}
\label{O3b_table}
\end{table*}

\section{Closing Remarks}

The precession parameter $\chi_p$ is a normalized measure of the maximum projection of spin misalignment onto the orbital plane between the two objects in a binary system. As such this parameter takes information from only one object, averaging over only the azimuthal angle difference $\Delta\Phi$, and does not account for the variation of the polar angles $\theta_1, \theta_2$ on the precession timescale $t_{pre} \propto \left(r/M\right)^{5/2}$. This inconsistency in averaging is resolved by the parameter $\langle \chi_p \rangle$, which averages over all the angular variations on the precession timescale and preserves the spin-misalignment information from both objects in the binary.\\

The generalized precession parameter $\langle \chi_p \rangle$ has now been implemented in the parameter estimation algorithm RIFT, and is now available in the latest release of RIFT at \url{https://github.com/oshaughn/research-projects-RIT}. With this implementation, both $\chi_p$ and $\langle \chi_p \rangle$ can be calculated for any set of intrinsic parameters and can therefore be used to generate posterior distributions from existing sample data. Such posteriors can be created and plotted directly from past and future posterior sample data using the \texttt{plot\_posterior\_corner.py} tool. For more information on using RIFT and associated functions, see \url{https://github.com/oshaughn/RIFT_tutorials}. It should be emphasized that the calculation of these parameters presented in this work were not sampled as part of the ILE/CIP process described in Sec.(\ref{riftreview}). An analysis of this sampling technique and its effect on parameter estimation of precessing systems is planned for future work, along with comparisons to alternative parameterizations of precession and the influence of higher-order waveform modes \cite{Thomas2021, Ramos-Buades2020}.\\

Additionally the precession characteristics for all 36 events from GWTC-3 have been reported. It was shown that there are ten events displaying high precession $\left(\langle \chi_p \rangle > 0.5\right)$, for which there are significant probabilities that their originating systems necessarily contain two misaligned spins. Further analysis of these events, in particular GW191109\_010717, is planned for a follow-up paper.

\section*{Acknowledgements}

This research has made use of data, software and/or web tools obtained from the Gravitational Wave Open Science Center (https://www.gw-openscience.org/ ), a service of LIGO Laboratory, the LIGO Scientific Collaboration and the Virgo Collaboration. LIGO Laboratory and Advanced LIGO are funded by the United States National Science Foundation (NSF) as well as the Science and Technology Facilities Council (STFC) of the United Kingdom, the Max-Planck-Society (MPS), and the State of Niedersachsen/Germany for support of the construction of Advanced LIGO and construction and operation of the GEO600 detector. Additional support for Advanced LIGO was provided by the Australian Research Council. Virgo is funded, through the European Gravitational Observatory (EGO), by the French Centre National de Recherche Scientifique (CNRS), the Italian Istituto Nazionale di Fisica Nucleare (INFN) and the Dutch Nikhef, with contributions by institutions from Belgium, Germany, Greece, Hungary, Ireland, Japan, Monaco, Poland, Portugal, Spain. This material is based upon work supported by the LIGO Laboratory which is a major facility fully funded by the NSF. The authors are grateful for computational resources provided by LIGO laboratory, supported by NSF Grants PHY-1841475, PHY-2110481, and PHY-2012057. Additional thanks to Richard U'dall for helpful discussions and comments. This paper has LIGO document number P2100434.

\centering
\noindent\rule{8cm}{0.4pt}

%----------------------------------------------------------------------------------------
%	REFERENCE LIST
%----------------------------------------------------------------------------------------
\section*{References \label{refs}}
\printbibliography[heading=none]

%----------------------------------------------------------------------------------------

\end{document}